\documentclass{aa}
\usepackage{graphicx}
\usepackage[round]{natbib}
\begin{document}

\title{Electric field screening by a proton counterflow
in the pulsar polar cap}

\author{Katsuaki Asano
        \inst{1}
        \and Fumio Takahara \inst{2}}

\institute{Division of Theoretical Astronomy, National Astronomical Observatory of Japan,
              2-21-1 Osawa Mitaka Tokyo, Japan\\
              \email{asano@th.nao.ac.jp}
         \and
           Department of Earth and Space Science,
            Osaka University, Toyonaka 560-0043, Japan\\
          \email{takahara@vega.ess.sci.osaka-u.ac.jp}}

\titlerunning{Electric Field Screening in the Pulsar Polar Cap}
\authorrunning{Asano and Takahara}

\abstract{
We propose a new mechanism to screen the electric field
in the pulsar polar cap. Previous studies have shown that 
if an electron beam from the stellar surface is accelerated 
to energies high enough to create electron-positron pairs, 
the required electric field parallel to the magnetic field 
lines is too strong to be screened out by the produced pairs.
We argue here that if non-relativistic protons are supplied 
from the magnetosphere to flow towards the stellar surface,
they can provide an anode to screen out such a strong 
electric field. Injected electron-positron pairs yield an 
asymmetry of the electrostatic potential around the screening 
point. The required pair creation rate in this model is 
consistent with the conventional models.
}

\maketitle

\section{INTRODUCTION}

\indent

A spinning magnetized neutron star provides huge electric
potential differences between different parts of its surface 
as a result of unipolar induction \citep{gol69}.
A part of the potential difference may be expended on an 
electric field along the magnetic field somewhere in the 
magnetosphere. Although a fully self-consistent model for the 
pulsar magnetosphere has not yet been constructed, several promising 
models have been proposed. Among them, the polar cap model 
\citep{stu71, rud75} assumes that an electric field 
$E_\parallel$ parallel to the magnetic field lines exists just
above the magnetic poles.
The electric field accelerates charged particles up to TeV energies,
and resultant curvature radiation from these particles produces copious 
electron-positron pairs through magnetic pair production.
These pairs may be sources for gamma-ray emission,
coherent radio emission, and the pulsar wind.

The localized potential drop is maintained by a pair of anode
and cathode regions.
In the cathode region the space charge density $\rho$ deviates
from the Goldreich-Julian (GJ) density 
negatively. 
On the other hand, $\rho$ must deviate positively for the anode.
Outside the accelerator the electric field will be screened.
In the polar cap model, especially for a space charge-limited flow model
\citep{faw77, sch78, aro79}, 
where electrons can freely escape from the stellar surface, 
i.e., $E_\parallel = 0$ on the stellar surface,
the formation mechanism of a static anode-cathode pair 
that can sustain enough potential drop 
for pair production is a long-standing issue. 
Since the current flows steadily along the magnetic field lines, 
the charge density is determined by the magnitude of the current 
and the field geometry with suitable boundary conditions. 
Basic considerations on this problem for space charge-limited flows 
are given in \citet{shi97}.
The mechanism of the electric field screening, i.e., a way
to provide an anode, may be provided by pair polarization. 
Although most papers take it for granted that copious pair production
can instantly screen the field, recently Shibata et al.
(1998, 2002: hereafter SMT98 and SMT02, respectively) have 
casted doubt on this issue; electric field screening is not as 
easy as most researchers consider conventionally.

SMT98 and SMT02 investigated the screening of electric fields
in the pair production region. 
They found that the required thickness of the screening layer is 
restricted to be as small as the braking distance 
$l_{\rm E}=m_{\rm e} c^2/|e E_\parallel|$, 
for which decelerating particles become non-relativistic,
where $m_{\rm e}$ and $-e$ are the electron mass and charge, respectively.
If the above condition does not hold,
too many positrons are reflected back and destroy the negative 
charge region (cathode).
In order to screen the electric field consistently,
a huge number of pairs should be injected within a region as small  
as $l_{\rm E}$.
The required pair multiplication factor per primary particle
is enormously large and cannot be realized in the conventional pair
creation models. Thus, some other ingredients are required for
the electric field screening.

In relation to this problem, \citet{lyu92} proposed a model 
with a positron counterflow and argued that the model
makes it possible to screen the electric field for any current density. 
He showed that the electric field on the stellar 
surface can be screened by the injection of electron-positron 
pairs plus a positron counterflow from the magnetosphere 
for any current density. In his model the positron counterflow is 
introduced to adjust the current desnity in the polar cap and 
does not play a critical role as an anode formation.
Most importantly, his model is different from the space charge-limited 
flow considered here and cannot be applied to our problem. 
As SMT98 and SMT02 have shown, the way to screen
the electric field at a reasonable altitude is unknown
for space charge-limited flows 
which we believe to be the most suitable model because the work function of 
the matter on the neutron surface is considered to be small enough.

In previous studies of the screening,
pairs were assumed to accelerate or decelerate
along the 0-th order trajectories determined by $E_\parallel$.
However, if an electrostatic (longitudinal) instability
occurs, the excited waves may produce effective friction which
will change the charge polarization process.
\citet{asa04} numerically investigated an electrostatic instability of 
electron-positron pairs injected in an external electric field.
However, they found that the density of electron-positron pairs
in the standard pulsar model is too low to 
induce the instability needed to provide strong friction.
This means that each particle moves stably along the trajectory 
determined by the external electric field since 
collective forces from other particles are negligible.

In this paper, we present a new model with a proton counterflow 
to solve the screening problem in the polar cap model with 
pair injection. Since the proton mass is 
much larger than the electron mass, its inertia effect is 
much stronger and can provide an anode structure more easily 
if non-relativistic protons are supplied from the outer 
magnetosphere to flow towards the polar caps. 
Our purpose is to point out a possibility
that the proton flow can explain the structure of the electric field
in the polar cap.
Therefore, our model is highly idealized and simplified.
Detailed investigations of the situation in our model 
are beyond the scope of the present paper. 
In Sect. 2 we briefly summarize the space charge-limited flow model
in the pulsar polar cap to give a basic framework for the present model.
In Sect. 3 we consider the model with proton counterflow and 
demonstrate the effectiveness of the screening
by the proton counterflow.
Sect. 4 is devoted to a summary and discussion.

\section{THE POTENTIAL FOR SPACE CHARGE LIMITED FLOW}

Before we present the model with a proton counterflow, 
we summarize the basic features of the behaviour of
the electric potential in the pulsar polar cap to make clear 
the fundamental points of the problem.
The notations are the same as in SMT98 and SMT02
in which the one-dimensional (1D) approximation is used.
We assume that $\mathbf{\Omega}_* \cdot \mathbf{B}>0$,
where $\mathbf{\Omega}_*$ and $\mathbf{B}$ are the angular velocity
of the star and magnetic field, respectively.
Although the current density distribution
is most likely to be determined by the global dynamics
in the magnetosphere, we will investigate a local region
just above the stellar pole.
We therefore treat the current density as an adjustable
free parameter in our paper, as assumed in SMT98 and SMT02.
A steady electron beam flows from the stellar surface
and is accelerated by a negative electric field parallel to
a magnetic field line with zero electric field on the stellar surface.
Beyond a certain altitude called the pair production front (PPF),
electron-positron pairs are assumed to be injected continuously into space.
In this case a part of the injected positrons
returns towards the stellar surface.
Here, we consider mainly the behavior of the electric potential
inside the PPF while that outside the PPF was treated in SMT02.

Since the current due to the primary electron beam flowing outward
and that due to positrons returning towards the stellar surface
flow inward along the magnetic field, both currents are negative.
Inside the PPF the difference in the charge density determined by these 
currents and the GJ density gives the effective charge density.
We define the GJ current density as
%%%%%%%%%%%%%%%%%%%%%%%
\begin{eqnarray}
J_0 \equiv -\frac{\mathbf{\Omega}_* \cdot \mathbf{B}}{2 \pi}.
\end{eqnarray}
%%%%%%%%%%%%%%%%%%%%%%%
We normalize the potential $\Phi$, length along the magnetic field line
$s$, GJ current density $J_0$, current density due to the primary 
electron beam $J$, and current density due to the returning positrons 
$J_+$ as
%%%%%%%%%%%%%%%%%%%%%%%
\begin{eqnarray}
\phi &\equiv& \frac{e \Phi}{m_{\rm e} c^2}, \\
l &\equiv& \frac{\omega_{\rm p} s}{c}, \\
j_0 &\equiv& \frac{J_0}{J_B}, \\
j &\equiv& \frac{J}{J_B}, \\
j_+ &\equiv& \frac{J_+}{J_B},
\end{eqnarray}
%%%%%%%%%%%%%%%%%%%%%%%
where
%%%%%%%%%%%%%%%%%%%%%%%
\begin{eqnarray}
J_B &\equiv& -\frac{\Omega_* B}{2 \pi} <0, \\
\omega_{\rm p}^2 &\equiv& \frac{4 \pi e |J_B|}{m_{\rm e} c}.
\end{eqnarray}
%%%%%%%%%%%%%%%%%%%%%%%
The current densities $J$, $J_+$, $J_0$, and $J_B$ are all negative,
so that $j_0$, $j$, $j_+$ are all positive, and by definition 
$j_0 \le 1$.
The equation of continuity ensures that $j \propto J/B$ and 
$j_+ \propto J_+/B$ are constant along the magnetic field line.
On the other hand, $j_0 \propto B_z/B$, where $B_z$
is the component of the magnetic field along the rotation axis, 
may change along the field line.
When the magnetic field line curves away from the rotation axis,
$j_0$ decreases with increasing $l$.
Conversely, when the field line curves toward the rotation axis,
$j_0$ increases with $l$.
Adopting the 1D approximation,
the non-dimensional Poisson equation becomes
%%%%%%%%%%%%%%%%%%%%%%%
\begin{eqnarray}
\frac{d^2 \phi}{d l^2}=\frac{j}{\beta}+\frac{j_+}{\beta_+}-j_0
\simeq \frac{j}{\beta}-j_+-j_0,
\label{eq}
\end{eqnarray}
%%%%%%%%%%%%%%%%%%%%%%%
where $\beta=v/c>0$ and $\beta_+=v_+/c<0$
are velocities of the primary electron beam and the returning positrons
normalized by the light velocity, respectively.
Since positrons are ultra-relativistic inside the PPF,
we have approximated as $\beta_+ \simeq -1$ here.

The boundary conditions are given on the stellar surface 
($l=0$) as follows; $\phi=0$, $d \phi/dl=0$, and the electron 
beam is supplied with $\beta=0$. In this case,
the first term due to the electron beam,
on the right hand side of Eq. (\ref{eq}),
dominates the other terms on the stellar surface.
The non-relativistic electron beam provides a huge negative 
charge density at $l=0$.
As a result, just above the stellar surface
$d^2 \phi/dl^2 >0$, which means that the potential curve is 
downward convex, irrespective of $j_+$ and $j_0$.
As long as the first term $j/\beta$ dominates,
$\phi$ increases with increasing $l$ and
the electric field becomes negative.
The electric field accelerates the electron beam,
and $\beta$ becomes $\simeq 1$ at a certain altitude.
Hereafter, we represent
the scale of the transition from $\beta \sim 0$ to $\sim 1$
as $l_{\rm nr}$;
for $l>l_{\rm nr}$, we write $\beta \simeq 1$.
For $l>l_{\rm nr}$, the Poisson equation
is approximated as
%%%%%%%%%%%%%%%%%%%%%%%
\begin{eqnarray}
\frac{d^2 \phi}{d l^2}=- \rho_{\rm in},
\label{eq2}
\end{eqnarray}
%%%%%%%%%%%%%%%%%%%%%
where
%%%%%%%%%%%%%%%%%%%%%%%
\begin{eqnarray}
\rho_{\rm in} \equiv -j+j_+ + j_0,
\label{rhoin}
\end{eqnarray}
%%%%%%%%%%%%%%%%%%%%%%%
is the effective charge density inside the PPF except for $0<l<l_{\rm nr}$.
In general, the scale $l_{\rm nr}$ is
much smaller than the length from the stellar surface to the PPF.
In the greater part inside the PPF, the potential obeys
Eq. (\ref{eq2}).

When we assume that $j_0$ is constant neglecting effects of 
field line curvature, $\rho_{\rm in}$ is constant.
In this case, the behaviour of the potential is determined
only by $\rho_{\rm in}$ and the situation is classified into the 
two cases: $\rho_{\rm in}<0$ and $\rho_{\rm in}>0$.
Hereafter we call the case of $\rho_{\rm in}<0$ ($j>j_+ + j_0$) 
Super-GJ and that of $\rho_{\rm in}>0$ ($j<j_+ + j_0$) Sub-GJ.

\subsection{Sub-GJ}

First, we discuss Sub-GJ. In the region inside $l_{\rm nr}$,
the non-dimensional electric field $E=-d \phi/dl<0$
grows with $l$ as was mentioned before.
For Sub-GJ, $d^2 \phi/dl^2$ for $l>l_{\rm nr}$ is negative,
so that the potential curve is upward convex.
In this region the absolute value of $E$ declines with $l$
as schematically shown in Fig. 1.
The electric field can be screened at a certain distance
even without pair formation. 
In the approximation of constant $\rho_{\rm in}$,
the solution of Eq. (\ref{eq2})
is obtained as
%%%%%%%%%%%%%%%%%%%%%%%
\begin{eqnarray}
\phi&=&- \frac{\rho_{\rm in}}{2} l^2+E_0 l,
\label{po} \\
E&=&-\frac{d \phi}{d l}=-E_0+ \rho_{\rm in} l,
\label{e1}
\end{eqnarray}
%%%%%%%%%%%%%%%%%%%%%%%
where $E_0>0$ represents the magnitude of the electric field at 
$l=0$. Strictly speaking, for $l<l_{\rm nr}$ the above solution 
is not correct and the electric field grows from 0 at 
$l=0$ to $-E_0$ inside $l_{\rm nr}$. The length $l_{\rm nr}$
is negligibly small when considering the region $l>l_{\rm nr}$.
From Eq. (\ref{e1}), the screening altitude, where the electric field
becomes zero again, is approximated as
%%%%%%%%%%%%%%%%%%%%%%%
\begin{eqnarray}
l_{\rm scr}=\frac{E_0}{\rho_{\rm in}}.
\label{koudo1}
\end{eqnarray}
%%%%%%%%%%%%%%%%%%%%%%%

\begin{figure}
\centering
   \includegraphics[width=8cm]{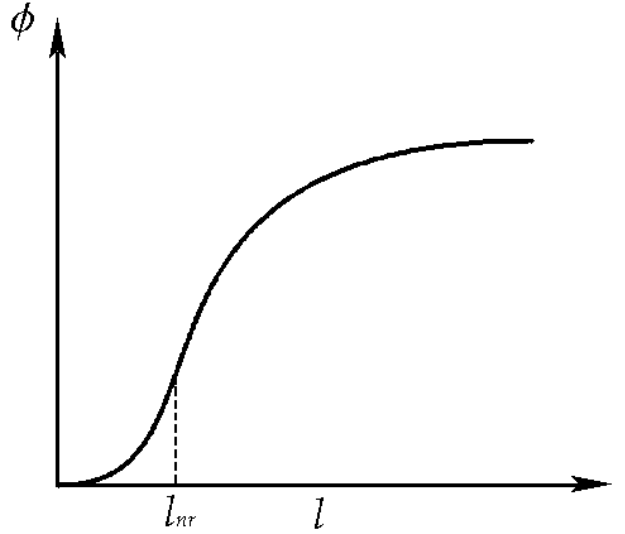}
\caption{Potential for Sub-GJ and constant $j_0$.
The position $l=l_{\rm nr}$ is the inflection point.
For $l<l_{\rm nr}$, the effective charge density is negative,
owing to the non-relativistic electron beam.
For $l>l_{\rm nr}$, the effective charge density
becomes $\rho_{\rm in}>0$.}
\end{figure}

If there is no pair injection, the electric field is not screened but 
changes sign at $l_{\rm scr}$ so that the electrons are decelerated 
outside $l_{\rm scr}$. Thus, the model becomes inconsistent. 
Let us further assume that around $l_{\rm scr}$ electron-positron 
pairs are injected and the electric field is screened.
We represent the Lorentz factor of the beam
at $l=l_{\rm scr}$ as $\Gamma(=\phi(l_{\rm scr})+1)=10^6 \Gamma_6$.
Using $\Gamma$, the electric field at $l=0$
is written by
%%%%%%%%%%%%%%%%%%%%%%%
\begin{eqnarray}
E_0 \simeq \sqrt{2 \rho_{\rm in} \Gamma}.
\label{denba}
\end{eqnarray}
%%%%%%%%%%%%%%%%%%%%%%%
Substituting this into Eq. (\ref{koudo1}),
the altitude of the screening is obtained as
%%%%%%%%%%%%%%%%%%%%%%%
\begin{eqnarray}
s_{\rm scr}&=&\frac{c l_{\rm scr}}{\omega_{\rm p}} =\sqrt{\frac{2 \Gamma}{\rho_{\rm in}}}
\frac{c}{\omega_{\rm p}}, \\
&=& 5 \times 10^4 \Gamma_6^{1/2}
\rho_{{\rm in},-3}^{-1/2}
T_{0.3}^{1/2}
B_{12}^{-1/2}
\quad {\rm cm},
\label{koudo2}
\end{eqnarray}
%%%%%%%%%%%%%%%%%%%%%%%
where $\rho_{{\rm in},-3} = \rho_{\rm in}/10^{-3}$, $B_{12}=B/10^{12}$ G
and $T=0.3 T_{0.3}$ s is the rotation period of pulsar.

To estimate $l_{\rm scr}$, we must estimate plausible 
ranges of $E_0$, $\rho_{\rm in}$ and $\Gamma$. 
The value of $\Gamma$ is considered to be $10^6\sim 10^7$ to 
obtain copious pair production. To constrain $E_0$ and 
$\rho_{\rm in}$, let us consider the behaviour of the potential
just above the stellar surface ($l<l_{\rm nr}$).
Multiplying Eq. (\ref{eq}) by $E$
and integrating it from the stellar surface
to an arbitrary height,
%%%%%%%%%%%%%%%%%%%%%%%
\begin{eqnarray}
-\frac{1}{2} \left[ E^2 \right]_{l=0}^{l=l}=\int_0^{\phi}
d \phi \left(j_0-\frac{j}{\beta}+j_+ \right),
\end{eqnarray}
%%%%%%%%%%%%%%%%%%%%%%%
where we use the relation $E dl=-d \phi$.
The above condition corresponds to the Gauss law; the column charge 
density between 
$l$ and the stellar surface is equal to $E(l)$.
Then we obtain
%%%%%%%%%%%%%%%%%%%%%%%
\begin{eqnarray}
-\frac{1}{2} E^2=(j_0+j_+) \phi-j \sqrt{\phi^2+2 \phi}.
\label{conc}
\end{eqnarray}
%%%%%%%%%%%%%%%%%%%%%%%
For Sub-GJ, $|E|$ takes the maximum value at $\phi=\phi_{\rm m}$
which is given by 
%%%%%%%%%%%%%%%%%%%%%%%
\begin{eqnarray}
\phi_{\rm m}^2+2 \phi_{\rm m}=\frac{A^2}{1-A^2},
\end{eqnarray}
%%%%%%%%%%%%%%%%%%%%%%%
where $A \equiv j/(j_0+j_+)<1$.
From Eq. (\ref{conc}), we obtain the maximum electric field as
%%%%%%%%%%%%%%%%%%%%%%%
\begin{eqnarray}
\frac{1}{2} E_{\rm max}^2 =(1-\sqrt{\mathstrut 1-A^2})(j_0+j_+)<j_0+j_+,
\label{emax}
\end{eqnarray}
%%%%%%%%%%%%%%%%%%%%%%%
We consider that this electric field gives an estimate of 
$E_0$ in Eq. (\ref{denba}). Then we have 
%%%%%%%%%%%%%%%%%%%%%%%
\begin{eqnarray}
\Gamma \simeq \frac{E_{\rm max}^2}{2 \rho_{\rm in}}
< \frac{j_0+j_+}{\rho_{\rm in}}.
\label{rg}
\end{eqnarray}
%%%%%%%%%%%%%%%%%%%%%%%
Since $\Gamma$ is very large, $\rho_{\rm in}$ should be very small 
as long as $j_0$ and $j_+$ are of order of unity. 

In order to complete the electric field screening,
the effective charge density should be zero outside $l_{\rm scr}$
which is identified with the PPF. Outside the PPF, excess pair electrons 
provide the normalized current $j_-$ which is equal to the 
current $j_+$ due to returning positrons inside the PPF.
Using the approximation that both electrons and positrons are relativistic,
the effective charge density outside the PPF is given by
%%%%%%%%%%%%%%%%%%%%%%%
\begin{eqnarray}
\rho_{\rm out} \equiv -j-j_+ + j_0=0.
\label{out}
\end{eqnarray}
%%%%%%%%%%%%%%%%%%%%%%%
Thus, $j_+=j_0-j$ should be satisfied for self-consistency.
The value $j_+$ to be positive requires $j<j_0$, 
which is equivalent to the Sub-GJ condition without pair injection.
Then the constraint of a very small $\rho_{\rm in}=2(j_0-j)=2j_+$
implies that the number of returning positrons is very small and that 
$E_0$ is of order of unity (see Eq. (\ref{emax})). If we assume $\Gamma \sim 10^6$,
$\rho_{\rm in} \sim 10^{-6}$, which implies a high screening 
height of $s_{\rm scr} \sim 10^6$ cm
from Eq. (\ref{koudo2}).
This altitude is much higher than conventionally supposed and 
actually, geometrical effects will play a role either in screening 
or in changing the flow into a Super-GJ flow.  

\subsection{Super-GJ}

\begin{figure}
\centering
   \includegraphics[width=8cm]{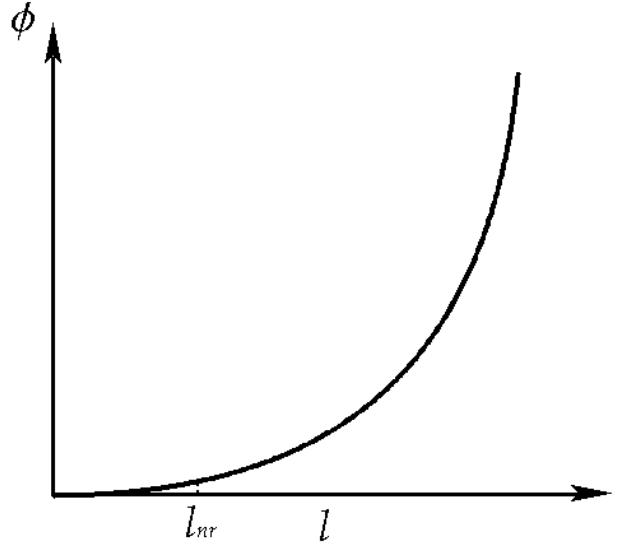}
\caption{Potential for Super-GJ and constant $j_0$.
The effective charge density is negative everywhere
independently of the velocity of the electron beam.}
\end{figure}

For Super-GJ, since $\rho_{\rm in}<0$ and $d^2 \phi/dl^2 >0$ 
the potential curve is 
downward convex everywhere (see Fig. 2),
The potential and the amplitude of the electric field 
steadily increase with $l$.
The electric field is given by Eq. (\ref{e1}) and becomes 
%%%%%%%%%%%%%%%%%%%%%%%
\begin{eqnarray}
E \simeq \rho_{\rm in} l,
\label{e2}
\end{eqnarray}
%%%%%%%%%%%%%%%%%%%%%%%
for large $l$ by neglecting $E_0$. 
If we conventionally assume that pair production can screen 
the electric field and if we also assume that the screening 
height $l_{\rm scr}$ roughly corresponds to the PPF,  
the screening height turns out to be given by Eq. 
(\ref{koudo2}),
when $\rho_{\rm in}$ is replaced by $|\rho_{\rm in}|$.
The non-dimensional electric field around the PPF is obtained as
%%%%%%%%%%%%%%%%%%%%%%%
\begin{eqnarray}
E_{\rm scr} \simeq -\sqrt{2 |\rho_{\rm in}| \Gamma}=-44.7 \times
|\rho_{{\rm in},-3}|^{1/2}
\Gamma_6^{1/2}.
\label{denba2}
\end{eqnarray}
%%%%%%%%%%%%%%%%%%%%%%%
If $|\rho_{\rm in}|$ is large, $E_{\rm scr}$ becomes large 
around the PPF, while the screening height becomes low.
As was shown in SMT02, pair polarization cannot screen the field
when $E_{\rm scr}$ is more than a few. If $E_{\rm scr}$ 
is lower than a few, the situation is the same as for Sub-GJ, i.e., 
geometrical effects play a role without pair production. 
Another problem of Super-GJ is the following.
For Super-GJ, the condition that the effective charge density 
outside the PPF vanishes becomes 
\begin{equation}
\rho_{\rm out}=-j-j_++j_0=\rho_{\rm in}-2j_+=0,
\end{equation}
where all the components are assumed to be relativistic.
This condition is never satisfied for positive $j_+$ within 
the 1D approximation.
In order to achieve screening for Super-GJ,
some components should be non-relativistic.
The existence of the anode implies
that the effective charge is positive at the point where $E=0$.
Therefore the effective charge density should decrease outside this point.
As a result, the models may require that a space charge density wave
continues outside the PPF as was shown in SMT98.
In this wave, 
the speeds of electrons and positrons
go and return repeatedly between a non-relativistic speed and the light speed.
The average effective charge density outside of the PPF is zero,
while the local charge density oscillates spatially.

\section{THE PROTON COUNTERFLOW MODEL}

Our proton counterflow model is motivated by the difficulties 
discussed in the previous section and 
provides a mechanism for realizing relevant $E_{\rm scr}$ and 
$l_{\rm scr}$ for a reasonable $\Gamma$.
As was shown in SMT98 and SMT02 and discussed in the previous section,
it is difficult to screen the electric field
in the system composed of the electron beam and injected pairs only.
For Super-GJ, we need to screen an electric field that is as strong as
given by Eq. (\ref{denba2}).
In this section we show that non-relativistic protons flowing backward
can screen the electric field owing to the large proton mass.

\subsection{Origin of the Proton Counterflow}

To resolve the screening problem, we add a proton counterflow 
as a minimal alteration to the standard model. The existence and 
origin of the proton counterflow is highly speculative and it is 
difficult to give definite mechanisms at the present stage. 
Although it is highly unlikely that protons flow towards the polar 
cap from outside the light cylinder, the proton counterflow can 
be realized if protons exist in the magnetosphere. Considering 
that we still do not well understand the fundamental issues of pulsar 
physics such as the current closure problem, we should be free from 
any kind of prejudices. 

There are some candidates for the origin of the proton 
counterflow. One of the most plausible candidates for the source
is protons existing in the co-rotating magnetosphere. In the 0-th 
approximation such protons are confined to closed magnetic lines 
so that the protons do not flow into the polar cap region.
However, if transport phenomena like the anomalous diffusion 
found in laboratory plasmas occur, the protons can flow into the 
polar cap region across magnetic field lines.
Electrostatic (longitudinal) drift-waves are considered to be the 
most dominant diffusion process in tokamaks \citep{lie85},
while vast amounts of theoretical efforts have been devoted to 
the understanding of the anomalous diffusion.

A density gradient of particles is likely to exist on the 
boundary between open and closed field lines. The drift motion due to 
the density gradient occurs along the direction (hereafter $y$-axis) 
perpendicular to both field lines and the density gradient ($x$-axis). 
Electrostatic drift-waves are excited accompanying the drift motion, 
so that the induced electric field along the $y$-axis induces 
$\mathbf{E} \times \mathbf{B}$ drift and particles diffuse along 
the $x$-axis, which will provide the source of the proton counterflow.
The diffusion coefficient for this process may be obtained from 
numerical simulations. 

The source of the anomalous diffusion is not limited to the 
electrostatic drift-waves due to the density gradient.
The extremely high brightness temperature of pulsar radio emission
requires a coherent source for this radiation and excitation of 
strong plasma waves in the open field line region. These waves 
may affect the corotating magnetosphere to induce proton diffusion.

As an example, let us roughly estimate the amount of protons
that diffuse from the co-rotating magnetosphere. 
Assuming that the diffusion occurs around the ``outer region''
(radius $R_{\rm O} \sim R_{\rm L}/2$, where 
$R_{\rm L} \equiv c/\Omega_*$ is the radius of the light cylinder).
Since the magnetic field is relatively weak
(typically $10^3$-$10^6$ G), the drift velocity $v_\bot \sim cE/B$ 
due to the excited electric field can be large enough.
Here, we assume that the scale of the proton diffusion region 
$\Delta R$ is much smaller than $R_{\rm O}$.
We denote $\Delta R=f R_{\rm O}$ ($f \ll 1$) hereafter.
Then, the number of protons escaping from the co-rotating
magnetosphere is written as
%%%%%%%%%%%%%%%%%%%%%%%
\begin{eqnarray}
2 \pi R_{\rm O} \Delta R \delta n_{\rm O} v_\bot=
\pi f (c/\Omega_*)^2 \delta n_{\rm O} v_\bot/2,
\label{diffuse}
\end{eqnarray}
%%%%%%%%%%%%%%%%%%%%%%%
where $\delta n_{\rm O}$ is the proton number density
resonating with the excited waves.

As will be shown below, our model requires that the proton current 
density on the polar cap is of the order of the GJ-current density.
The typical size of the polar cap region $R_{\rm pol}$ on the 
stellar surface is $\sim 10^4$ cm.
In our model the number of protons flowing into the polar cap region 
per unit time is
%%%%%%%%%%%%%%%%%%%%%%%
\begin{eqnarray}
\dot{N}_{\rm p}
\sim \frac{J_0}{e} R_{\rm pol}^2 
\sim \frac{\Omega_* B_* R_{\rm pol}^2}{2 \pi e},
\end{eqnarray}
%%%%%%%%%%%%%%%%%%%%%%%
where $B_*$ is the magnetic field on the stellar pole.
The supply rate given by Eq. (\ref{diffuse}) is larger 
than the required rate if 
%%%%%%%%%%%%%%%%%%%%%%%
\begin{eqnarray}
\delta n_{\rm O}
&>&\frac{B_* R_{\rm pol}^2 \Omega_*^3}{\pi^2 c^2 e f v_\bot} \\
&\sim& 700 
\times \nonumber \\
&& T_{0.3}^{-3} B_{12} 
\left( \frac{R_{\rm pol}}{10^4 {\rm cm}} \right)^2
\left( \frac{f}{0.1} \right)^{-1}
\left( \frac{v_\bot}{0.1 c} \right)^{-1} {\rm cm^{-3}}.
\label{nout}
\end{eqnarray}
%%%%%%%%%%%%%%%%%%%%%%%
This value seems very small for the atmosphere around such a high 
density celestial object, and it is less than the Goldreich-Julian 
density at $R_{\rm O}$, though there are large uncertainties for $v_\bot$ 
and $f$.
In some atmosphere models \citep{ho01,zan01} that are consistent with
observed proton-cyclotron resonance in a soft gamma repeater
or an anomalous X-ray pulsar \citep{ibr03,rea03}, the proton density 
is assumed to be larger than $10^{20} {\rm cm^{-3}}$.
Although the situation here is quite different from such models,
the proton density in the magnetosphere may be much higher than
the value of Eq. (\ref{nout}). Therefore, we consider that it 
is not so unlikely that a fraction of the protons diffuse from the 
co-rotating magnetosphere flow into the polar cap region, and that the 
global dynamics in the magnetosphere adjusts the current density 
distribution to screen the electric field.

Further detailed discussion of the diffusion depends on
the wave excitation mechanism, the model of global current flows,
the model of the atmosphere, and the convection in the 
co-rotating magnetosphere etc., which are not well understood. 
We will not discuss those problems in this paper. There may be  
mechanisms for the diffusion from the co-rotating magnetosphere 
other than described here and we do not stick to a specific mechanism.
The existence of protons may be connect with the current closure problem
and the source and role of proton flows are interesting subjects in 
connection with all the problems in pulsar physics.

\subsection{Overview of the Model}

The schematic picture for the proton counterflow model is 
depicted in Fig. 3.
As shown in this figure, our model incorporates a bulge of the potential,
which is divided into the following four regions:
region I ($E<0$, effective charge density $\rho_{\rm eff}<0$),
II ($E<0$, $\rho_{\rm eff}>0$), III ($E>0$, $\rho_{\rm eff}>0$),
and IV ($E>0$, $\rho_{\rm eff}<0$).
The potential increases from the stellar 
surface through regions I and II and peaks at the boundary 
between regions II and III.
At the peak of the potential the electric field becomes zero.
Then the potential decreases and the electric field
becomes positive in regions III and IV.
Our model requires that the electric field vanishes
again in region IV.
The potential curve in region IV is downwards convex and 
reaches the bottom again
(we call this point ``valley'', hereafter).
In our model we assume that pair electrons do not return,
so that the potential drop should be smaller than
the electron energy at injection.
As was mentioned in Sect. 2.2, the potential continues to
oscillate outside of the valley.
The potential curve shown in Fig. 3 is possible if
the proton counterflow exists,
as will be explained below.

\begin{figure}
\centering
   \includegraphics[width=8cm]{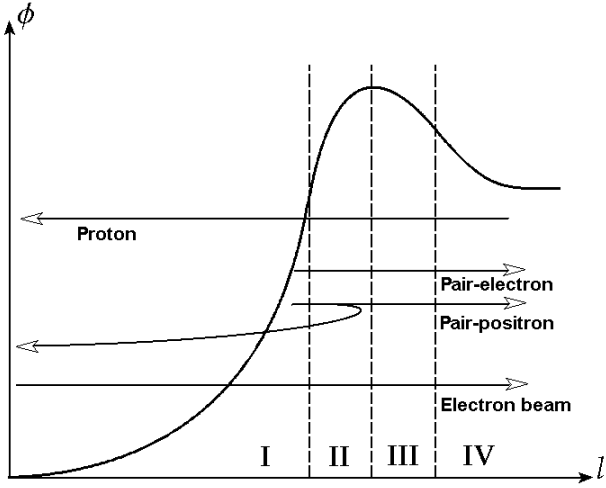}
\caption{Schematic picture for the proton counterflow model.}
\end{figure}

We assume that the PPF is located in region I or II.
Positrons in regions I and II are decelerated,
and a part of the positrons returns towards the stellar surface.
Therefore, in most of region I
there are three relativistic flows: the electron beam coming 
from the stellar surface,
the returning positron flow, and the proton counterflow.
In this region the effective charge density determined
from these flows is assumed to be negative (Super-GJ).
As is discussed in Sect. 2.2,
the amplitudes of the potential and electric field grow with $l$ in region I.
As the potential increases with $l$, protons become non-relativistic
and their speed decreases with $l$.
As a result, the positive charge density due to protons increases.
In regions II and III the effective charge density is positive
and the potential curve is upwards convex,
because of the contribution of the non-relativistic protons.
In region IV the effective charge density once again becomes negative,
because the speed of protons increases with $l$.

The bulge of the potential is formed by the positive charge density
due to the proton counterflow.
Electron-positron pairs cause an asymmetry between
region I-II and region III-IV.
If there is no pair injection,
region I-II and region III-IV are symmetric
and the valley becomes $\phi=0$. In this case, 
the potential becomes periodic within the 1D approximation.
When pairs are injected, as is required for pulsar action,
part of the positrons cannot cross the potential peak and 
return towards the stellar surface.
On the other hand, all injected electrons
cross the peak and flow into region III.
Thus, there is an asymmetry of the charge density
between regions I-II and III-IV.
In our model
the local effective charge density in region IV ($\sim -j_+$) is
much larger than that in region I ($|\rho_{\rm in}| \ll 1$).
The existence of excess negative charge 
due to pair electrons in region IV leads to
an immediate change of $d \phi/d l$ with $l$, and
will keep the valley shallow enough.

\subsection{Region of $E<0$}

In this subsection we discuss regions I and II.
Following SMT02, 
outside the PPF the potential satisfies
the Poisson equation,
%%%%%%%%%%%%%%%%%%%%%%%
\begin{eqnarray}
\frac{d^2 \phi}{d l^2}=j+\frac{j_{\rm p}}{\beta_{\rm p}}
-j_0-\bar{\rho}_+ - \bar{\rho}_-,
\label{poi}
\end{eqnarray}
%%%%%%%%%%%%%%%%%%%%%%%
where $j_{\rm p}>0$ and $\beta_{\rm p}<0$
are the normalized current density and velocity of the proton 
counterflow,
respectively, and $\bar{\rho}_+ \equiv c \rho_+/|J_B|$
and $\bar{\rho}_- \equiv c \rho_-/|J_B|$ are non-dimensional
charge densities of pair positrons and electrons, respectively.
Since pairs are injected for a finite length; their densities 
are calculated by integrating over the injection place, 
which will be done below.  

Let us set $\phi=0$ and $E=E_{\rm F}$ at the PPF.
At the peak of the potential ($\phi=\phi_{\rm M}$),
namely the boundary between regions II and III,
the electric field becomes zero.
Integrating the Poisson equation between the PPF and the peak 
as was done in Sect. 2.1,
we obtain
%%%%%%%%%%%%%%%%%%%%%%%
\begin{eqnarray}
-\frac{1}{2} E_{\rm F}^2=&&\int_0^{\phi_{\rm M}} d \phi (j-j_0)
-\int_0^{\phi_{\rm M}} d \phi (\bar{\rho}_+ + \bar{\rho}_-) \nonumber \\
&&-\int_0^{\phi_{\rm M}} d \phi \frac{j_{\rm p}}{|\beta_{\rm p}|}.
\label{3term}
\end{eqnarray}
%%%%%%%%%%%%%%%%%%%%%%%
The first term of the right hand side in Eq. (\ref{3term}) 
becomes
%%%%%%%%%%%%%%%%%%%%%%%
\begin{eqnarray}
\int_0^{\phi_{\rm M}} d \phi (j-j_0)=(j-j_0) \phi_{\rm M}.
\end{eqnarray}
%%%%%%%%%%%%%%%%%%%%%%%

To calculate the second term we follow the same procedure 
as SMT02. We introduce the multiplication factor $m(l)$
in order to take into account the contribution of injected 
electron-positron pairs.
The net pair flux 
produced between the PPF and a position $l$ is denoted by $m(l) j$.
Since $\phi$ and $m$ are both monotonic functions of $l$ 
in this region,
$m$ or $\phi$ can be used to designate the coordinate position.
The charge densities of pairs injected by a small flux element between
$m$ and $m+dm$ are given by
%%%%%%%%%%%%%%%%%%%%%%%
\begin{eqnarray}
d \bar{\rho}_\pm =\pm j \frac{dm}{|\beta_\pm|} =\pm \frac{j \gamma_\pm}
{\sqrt{\gamma_\pm^2-1}} dm,
\end{eqnarray}
%%%%%%%%%%%%%%%%%%%%%%%
where the Lorentz factors $\gamma_\pm$ are measured at the position
$\phi$, and are functions of the injection point 
denoted by $\phi_{\rm inj}(m)$,
i.e.,
%%%%%%%%%%%%%%%%%%%%%%%
\begin{eqnarray}
\gamma_-=\phi-\phi_{\rm inj}(m)+\bar{\gamma}
\end{eqnarray}
%%%%%%%%%%%%%%%%%%%%%%%
for electrons and
%%%%%%%%%%%%%%%%%%%%%%%
\begin{eqnarray}
\gamma_+=-\phi+\phi_{\rm inj}(m)+\bar{\gamma}
\end{eqnarray}
%%%%%%%%%%%%%%%%%%%%%%%
for positrons; $\bar{\gamma}$ is the Lorentz factor at injection,
and assumed to be constant for simplicity.

A detailed analytic calculation of the second term in Eq. 
(\ref{3term})
in general cases is shown in SMT02.
Here, for simplicity, we adopt a toy model
for the pair injection rate;
the pair creation rate is constant with respect to the
`$\phi$-coordinate', as $dm=\alpha_\phi d\phi$,
where $\alpha_\phi$ is constant.
While $\alpha_\phi$ is the pair multiplication factor created
in a unit potential drop $m_{\rm e} c^2/e$,
taking a constant $\alpha_\phi$ is a good approximation for a spatially
uniform pair injection model (see SMT02).
In this case the charge density of pair electrons is easily calculated as
%%%%%%%%%%%%%%%%%%%%%%%
\begin{eqnarray}
\bar{\rho}_-(\phi)&=&-j \alpha_\phi \int_0^\phi d \phi_{\rm inj}
\frac{1}{\beta_-} \nonumber \\
&=&j \alpha_\phi
\left(\sqrt{\mathstrut \bar{\gamma}^2-1}
-\sqrt{(\bar{\gamma}+\phi)^2-1}\right).
\end{eqnarray}
%%%%%%%%%%%%%%%%%%%%%%%

At a given position $\phi$, the Lorentz factor of positrons
injected at $\phi_{\rm inj}=\phi-\bar{\gamma}+1$
becomes $\gamma_+=1$.
These positrons turn around there and return toward the stellar surface.
For $\phi>\bar{\gamma}-1$,
positrons injected between $\phi-\bar{\gamma}+1 <\phi_{\rm inj}<\phi$ 
move forward at a position $\phi$. Positrons injected 
at $\phi_{\rm inj}<1+\phi_{\rm M}-\bar{\gamma}$
return before $\phi=\phi_{\rm M}$
and do not enter region III, while 
positrons injected at $\phi_{\rm inj} \ge 1+\phi_{\rm M}-\bar{\gamma}$
enter region III and do not return.
Therefore, at a given position $\phi$,
positrons moving backward were injected at
$\phi-\bar{\gamma}+1 <\phi_{\rm inj}<1+\phi_{\rm M}-\bar{\gamma}$.
Thus, the charge density of pair positrons at $\phi$ is given by 
%%%%%%%%%%%%%%%%%%%%%%%
\begin{eqnarray}
\bar{\rho}_+(\phi)=&&j \alpha_\phi \int_{\phi-\bar{\gamma}+1}^\phi 
d \phi_{\rm inj}
\frac{1}{|\beta_+|} \nonumber \\
&&+ j \alpha_\phi \int_{\phi-\bar{\gamma}+1}^{1+\phi_{\rm M}-\bar{\gamma}} 
d \phi_{\rm inj}
\frac{1}{|\beta_+|},
\end{eqnarray}
%%%%%%%%%%%%%%%%%%%%%%%
where the first term of the right hand side is due to the positrons 
moving forward,
and the second is due to the positrons moving backward.
For $\phi \le \bar{\gamma}-1$, we should change all the lower 
bounds of the integrals
to zero.
As a result, we obtain
%%%%%%%%%%%%%%%%%%%%%%%
\begin{eqnarray}
\bar{\rho}_+ = j \alpha_\phi \left\{
\begin{array}{l}
\biggl(\sqrt{\mathstrut \bar{\gamma}^2-1}
+\sqrt{(\phi_{\rm M}-\phi+1)^2-1}\biggr), \\
\qquad \qquad \qquad \qquad \qquad 
\mbox{for $\phi>\bar{\gamma}-1$,} \\
\biggl(\sqrt{\mathstrut \bar{\gamma}^2-1}+\sqrt{(\phi_{\rm M}-\phi+1)^2-1} \\
\qquad \qquad \qquad -2 \sqrt{(\bar{\gamma}-\phi)^2-1}\biggr), \\
\qquad \qquad \qquad \qquad \qquad 
\mbox{for $\phi \le \bar{\gamma}-1$.} \\
\end{array} \right.
\end{eqnarray}
%%%%%%%%%%%%%%%%%%%%%%%

The charge density of pairs is straightforwardly integrated 
to give the second term in Eq. (\ref{3term}) as
%%%%%%%%%%%%%%%%%%%%%%%
\begin{eqnarray}
\int_0^{\phi_{\rm M}} d \phi (\bar{\rho}_+ + \bar{\rho}_-)
=j \alpha_\phi \Bigl[ 
(2 \phi_{\rm M}-\frac{1}{2} \bar{\gamma}) \sqrt{\bar{\gamma}^2-1} \nonumber \\
-\frac{1}{2} (\phi_{\rm M}+\bar{\gamma}) 
\sqrt{(\phi_{\rm M}+\bar{\gamma})^2-1} \nonumber \\
+\frac{1}{2} (\phi_{\rm M}+1) \sqrt{(\phi_{\rm M}+1)^2-1} \nonumber \\
+\frac{1}{2} \log{\frac{G(\phi_{\rm M}+\bar{\gamma})G(\bar{\gamma})}
{G(\phi_{\rm M}+1)}}
\Bigr],
\label{pair}
\end{eqnarray}
%%%%%%%%%%%%%%%%%%%%%%%
where
%%%%%%%%%%%%%%%%%%%%%%%
\begin{eqnarray}
G(\phi)=\phi+\sqrt{\phi^2-1}.
\end{eqnarray}
%%%%%%%%%%%%%%%%%%%%%%%

The third term of Eq. (\ref{3term}) represents the contribution 
of protons. 
The Lorentz factor of protons is given by
%%%%%%%%%%%%%%%%%%%%%%%
\begin{eqnarray}
\gamma_{\rm p}=(\phi_{\rm M}-\phi) \delta+\gamma_{\rm m},
\end{eqnarray}
%%%%%%%%%%%%%%%%%%%%%%%
where $\delta \equiv m_{\rm e}/m_{\rm p}$  
and $\gamma_{\rm m}\simeq 1$ is the minimum Lorentz factor of the proton 
counterflow at $\phi=\phi_{\rm M}$.
The third term in Eq. (\ref{3term}) becomes
%%%%%%%%%%%%%%%%%%%%%%%
\begin{eqnarray}
\int_0^{\phi_{\rm M}} d \phi \frac{j_{\rm p}}{|\beta_{\rm p}|}=
\frac{j_{\rm p}}{\delta}(\sqrt{(\delta \phi_{\rm M}+\gamma_{\rm m})^2-1}-
\sqrt{\gamma_{\rm m}^2-1}).
\end{eqnarray}
%%%%%%%%%%%%%%%%%%%%%%%

Since we have calculated the righthand side of Eq. (\ref{3term}), 
we now investigate whether the proton counterflow can make $E_{\rm F}$ 
large enough. 
One constraint on the parameters is imposed by the flow inside the PPF, 
because we are treating Super-GJ with $\rho_{\rm in}<0$. 
Since all components are relativistic inside the PPF except very near 
the stellar surface,
the charge density is approximated as
%%%%%%%%%%%%%%%%%%%%%%%
\begin{eqnarray}
\rho_{\rm in} \equiv j_0-j+j_+ + j_{\rm p}<0,
\end{eqnarray}
%%%%%%%%%%%%%%%%%%%%%%%
where $j_+$ is the non-dimensional current density due to returning 
positrons inside the PPF. 
Inside the PPF, positrons injected at 
$0 <\phi_{\rm inj}<1+\phi_{\rm M}-\bar{\gamma}$
are returning.
The charge density of positrons inside the PPF ($\phi<0$) is obtained as
%%%%%%%%%%%%%%%%%%%%%%%
\begin{eqnarray}
\bar{\rho}_+&=&j \alpha_\phi
\left(\sqrt{(\phi_{\rm M}-\phi+1)^2-1}-\sqrt{(\bar{\gamma}-\phi)^2-1}\right).
\end{eqnarray}
%%%%%%%%%%%%%%%%%%%%%%%
By definition, we have the relation
%%%%%%%%%%%%%%%%%%%%%%%
\begin{eqnarray}
{j_+}=j\alpha_\phi(1+\phi_{\rm M}-\bar{\gamma}).
\end{eqnarray}
%%%%%%%%%%%%%%%%%%%%%%%
It is seen that inside the PPF the charge density of positrons
is approximated by $\bar{\rho}_+=j_+$ with high accuracy.

First of all, let us confirm that $E_{\rm F}$
should be smaller than the value in Eq. (\ref{denba2})
when there is no proton counterflow.
In this case
%%%%%%%%%%%%%%%%%%%%%%%
\begin{eqnarray}
E_{\rm F}^2/2
= \int_0^{\phi_{\rm M}} d \phi (\bar{\rho}_+ + \bar{\rho}_-)
-(j-j_0) \phi_{\rm M}.
\end{eqnarray}
%%%%%%%%%%%%%%%%%%%%%%%
In the limit of $\phi_{\rm M} \ge \bar{\gamma} \gg 1$ and
$\Delta \equiv \phi_{\rm M}-\bar{\gamma} \gg 1$, we obtain
%%%%%%%%%%%%%%%%%%%%%%%
\begin{eqnarray}
E_{\rm F}^2/2 &\simeq& 
j_+ (\phi_{\rm M}/\Delta+\bar{\gamma})-(j-j_0) \phi_{\rm M} \nonumber \\
&=&(j-j_0) (\phi_{\rm M}/\Delta-\Delta)+\rho_{\rm in}
(\phi_{\rm M}/\Delta+\bar{\gamma}).
\label{neg}
\end{eqnarray}
%%%%%%%%%%%%%%%%%%%%%%%
We should notice that $j-j_0>0$ and $\rho_{\rm in}<0$ for Super-GJ considered here.
The second term on the righthand side is always negative.
If $\Delta \sim \phi_{\rm M}$ (namely $\phi_{\rm M} \gg \bar{\gamma}$),
the first term $\simeq -\Delta (j-j_0)$ becomes negative too.
Therefore, positive $E_{\rm F}^2$ 
is possible only for $\Delta^2 \ll \phi_{\rm M}$.
As is understood from Eq. (\ref{neg}),
$E_{\rm F}^2$ cannot be larger than $\sim 10^3$,
for $\Delta \gg 1$, $j-j_0 \sim O(1)$
and $\phi_{\rm M} \sim \bar{\gamma}=10^2$--$10^3$.
If $\Delta$ is of the order of unity or less,
$E_{\rm F}^2$ can be large.
In this case, however, $\alpha_\phi=j_+/j (1+\Delta)$ becomes of the order of unity,
while $\alpha_\phi \ll 1$
in realistic situations  (see SMT02).
Thus, only $E_{\rm F}$ of the order of unity 
can be screened through pair polarization.

When a proton counterflow exists, the situation changes drastically.
The third term of Eq. (\ref{3term}) is easily calculated,
and we obtain
%%%%%%%%%%%%%%%%%%%%%%%
\begin{eqnarray}
E_{\rm F}^2/2 \simeq
\frac{j_{\rm p}}{\delta}
\sqrt{(\delta \phi_{\rm M}+\gamma_{\rm m})^2-1}
+j_+ (\phi_{\rm M}/\Delta+\bar{\gamma}) \nonumber \\
-(j-j_0)\phi_{\rm M},
\end{eqnarray}
%%%%%%%%%%%%%%%%%%%%%%%
where the first and second terms are due to the proton counterflow
and pairs, respectively.
As long as the proton flow is non-relativistic,
$\gamma_{\rm m}$ is not important for determining $E_{\rm F}$,
though the local charge density depends strongly on $\gamma_{\rm m}$.
In conventional pair creation models,
$\bar{\gamma}$ is assumed to be a few hundreds,
namely $\bar{\gamma} \sim O(1/\delta)$.
If $\phi_{\rm M} \gg 1/\delta$,
$E_{\rm F}^2/2 \simeq (\rho_{\rm in}-j_+) \phi_{\rm M}+j_+ \bar{\gamma}
= \rho_{\rm in} \phi_{\rm M}-j_+ \Delta<0$.
Too large a value of $\phi_{\rm M}$ leads to a large negative contribution
of the last term.
Therefore, $E_{\rm F}^2$ becomes positive and large only when 
$\phi_{\rm M}$, $1/\delta$ and $\bar{\gamma}$ are of the same order,
but $1/\delta$ is a few times larger than the others; in this case 
the contribution of the second and third terms becomes
$\sim -\Delta j_+ - j_{\rm p} \phi_{\rm M}$.
Even if $\Delta \sim \phi_{\rm M}$,
this contribution is as small as $\sim -O(j_{\rm p} \phi_{\rm M})$
for $j_+ \sim j_{\rm p}$.
The proton contribution is a few times $j_{\rm p}/\delta$
and can provide a large value of $E_{\rm F} \sim \sqrt{j_{\rm p}/\delta}$.
When  $j_{\rm p}$ is of the order of unity, $E_{\rm F}$ is a few times ten,
which is close to the value in Eq. (\ref{denba2}).
The small value of $\delta$ makes $E_{\rm F}$ large enough.
When $j_+ \gg j_{\rm p}$,
the contribution of $-\Delta j_+$ in the second and third terms
dominates the other terms.
Therefore, the current $j_+$ is at most comparable to $j_{\rm p}$.

\subsection{Region of $E>0$}

In this subsection we discuss regions III and IV, where 
the electric field becomes positive 
so that positrons are accelerated while electrons are decelerated.
The speed of incoming protons increases with $l$. 
There are two possibilities to complete the screening; 
one is to vanish charge density and the other is 
to obtain a small-amplitude charge density wave with vanishing 
average charge density, which was shown in SMT98. 
The first possibility is unlikely because this requires that
the effective charge should vanish suddenly at the point where $E=0$,
although the charge density (proton density)
tends to increase toward the peak ($\rho_{\rm eff}>0$) of the potential
and decrease toward the valley ($\rho_{\rm eff}<0$).
Thus, the second possibility is more natural. 
The oscillation amplitude should be small enough since 
if the amplitude in the potential is too large in this region,
injected electrons would return to region II and continue to 
oscillate around the potential peak without escaping to infinity.

In this subsection we discuss the condition
for producing the valley.
The Poisson equation was given in Eq. (\ref{poi}).
Part of the pairs injected in regions I and II
flow into regions III and IV.
The injection region for such electrons is $0<\phi<\phi_{\rm M}$ 
in regions I and II,
and for positrons it is 
$1+\phi_{\rm M}-\bar{\gamma}<\phi<\phi_{\rm M}$.
As for pairs injected in regions III and IV,
we assume $dm=-\alpha_\phi d\phi$,
because $\phi$ decreases with $l$ in regions III and IV,
Then we obtain
%%%%%%%%%%%%%%%%%%%%%%%
\begin{eqnarray}
\bar{\rho}_-=j \alpha_\phi
\bigl(2 \sqrt{(\bar{\gamma}+\phi-\phi_{\rm M})^2-1}
-\sqrt{\mathstrut \bar{\gamma}^2-1} \nonumber \\
-\sqrt{(\bar{\gamma}+\phi)^2-1} \bigr), \\
\bar{\rho}_+=j \alpha_\phi
\bigl(2 \sqrt{(\bar{\gamma}-\phi+\phi_{\rm M})^2-1}
-\sqrt{\mathstrut \bar{\gamma}^2-1}  \nonumber \\
-\sqrt{(\phi_{\rm M}-\phi+1)^2-1} \bigr).
\end{eqnarray}
%%%%%%%%%%%%%%%%%%%%%%%
Since our model requires that the oscillation amplitude of the potential
is small enough,
all electrons and positrons injected in regions III and IV
are relativistic.
Therefore, the charge densities of such electrons and positrons
cancel out,
and scarcely contribute to the total charge density at all.
The charge density due to pairs ($\bar{\rho}_- + \bar{\rho}_+$)
becomes $\sim -j_+$ in regions III and IV,
which is confirmed from the above equations.

In region III
the positive charge due to protons dominates over the negative charge density
due to pairs, because of the low speed of the proton counterflow.
As the speed of the proton flows increases with $l$,
the negative charge due to pairs dominates in region IV and the valley
will appear.
Further out
the spatial wave of the potential and charge density
will repeat; this will be discussed in the next subsection.
The condition to make the effective charge density negative
in region IV is
$j_{\rm p}/|\beta_{\rm p}|
< j_+ + (j-j_0)=2 j_+ + j_{\rm p} - \rho_{\rm in}$.
On the other hand, opposite inequality applies in region III
because of the positive effective charge density in region III. 
A small value of $\beta_{\rm p}$ in region III
satisfies the condition.

The condition in region IV is rewritten as $|\beta_{\rm p}|>j_{\rm p}/
(2 j_+ + j_{\rm p} - \rho_{\rm in})$.
In order to obtain a larger value of $E_{\rm F}$, a smaller
value of $j_+ < \sim j_{\rm p}$ is favorable,
while a finite value of $j_+$ is required to create the assymetry
in the potential.
From the above conditions,
$\beta_{\rm p}$ should vary significantly ($\sim O(1)$),
which requires that the typical amplitude of 
potential variation is of the order of $1/\delta$.
At the same time, pair electrons and positrons are assumed to flow 
outward continuously, which requires a fairly large value of 
$\bar{\gamma}> \sim 1/\delta$ to overcome the potential variation.

In the next subsection
we will confirm numerically whether the model can reproduce both a shallow valley
and a strong $E_{\rm F}$.

\subsection{Numerical Demonstration}

We here numerically demonstrate the screening by the proton counterflow model
in order to confirm the behaviour of the potential in regions III and IV
especially.
In this subsection we assume that
$j_0=1.0$, $\rho_{\rm in}=-0.001$, $\bar{\gamma}=500$, and
$\beta_{\rm m} \equiv -\sqrt{1-1/\gamma_{\rm m}^2}=-0.1$.
The other parameters are $j_{\rm p}$, $j_+$, and the potential
difference between the PPF and the peak, $\phi_{\rm M}$.
The physical quantities $j$ and $\alpha_\phi$
are determined by the above parameters.
We assume that the currents $j_{\rm p}$, $j_+$, and $j$ are all $O(1)$.
From Eq. (\ref{3term}) we can determine the electric field at the PPF, $E_{\rm F}$.
By setting $E=E_{\rm F}$ and $\phi=0$ at the PPF,
we can numerically solve the Poisson equation from the charge densities
we have obtained in this section.
In Table 1 we show some examples of our numerical results
that succeed in screening the electric field with subsequent charge density wave.
As discussed in Sect. 3.2, pair electrons return from
region III when $\bar{\gamma}$ is as small as $100$, because the model requires
that potential variation is of the order of $1/\delta$.
Since $\phi_{\rm M} > \bar{\gamma}$ (namely $j_+ > 0$) in our model
to yield the asymmetry of the potential,
a larger $\bar{\gamma}$ leads to a larger $\phi_{\rm M}$.
As mentioned in Sect. 3.1, $\phi_{\rm M}$ should be $\sim 1/\delta$.
This is the reason why we have chosen $\bar{\gamma}=500$.
When we adopt a larger value of $\bar{\gamma}$,
the numerical results show that $E_{\rm F}$ becomes smaller.
Fig. 4 shows the potential and charge densities
for one of our results, Case 1.

\begin{figure}
\centering
   \includegraphics[width=8cm]{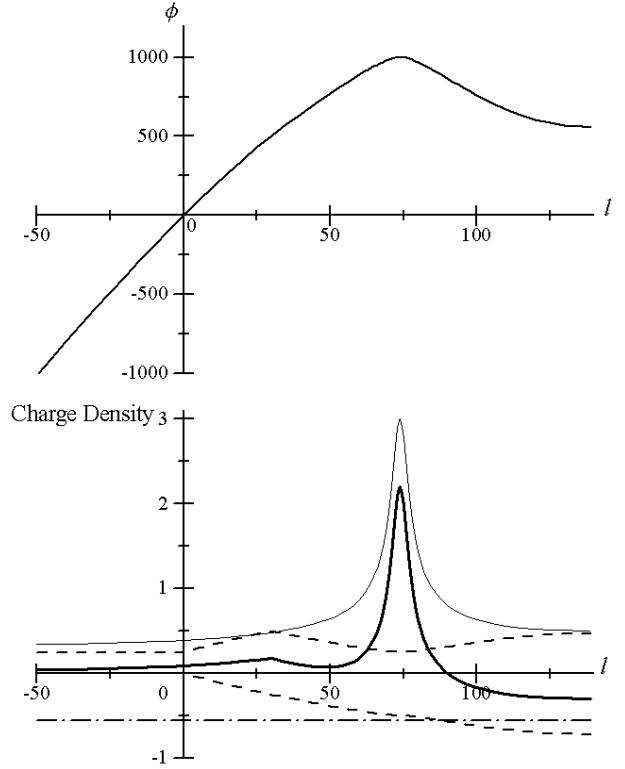}
\caption{Potential (upper) and charge densities (lower) for Case 1.
In the lower figure, the thick solid line is the effective total charge density.
Thin solid line, dashed lines, and dash-dotted line are
the charge densities due to protons ($-j_{\rm p}/|\beta_{\rm p}|$),
injected pairs ($\rho_-<0$ and $\rho_+>0$),
and the constant contribution ($-(j-j_0)=-0.55$), respectively.}
\end{figure}

Table 1 indicates that a smaller $\phi_{\rm M}$ 
and a larger $j_{\rm p}$ lead to a larger amplitude of $E_{\rm F}$,
as discussed in Sect. 3.1.
In all our examples the PPF ($\phi=0$ and $l=0$) is in region II.
The position where the electric field becomes strongest is
on the boundary of regions I and II.
Since the proton acceleration is slower than the electron acceleration,
the boundary is far inside the PPF.
For Case 1 the effective charge density becomes zero
at $l=-800$ where $\phi=-2 \times 10^4$.
At this point the amplitude of $E$ becomes maximum.
In Table 1 we also tabulate the maximum electric field $E_{\rm M}$
numerically obtained.
The value of $E_{\rm M}$ rather than $E_{\rm F}$
is the electric field screened in our model.
The results show that $E_{\rm M}$ is close to
the value in Eq. (\ref{denba2}).

%%%%%%%%%%%%%%%%%%%%%%%%%%%%%%%%%%%%%%%%%%%%%%%%%%%%%%%%
\begin{table*}
\centering
\begin{minipage}{100mm}
\caption{Model parameters and numerical results for $E_{\rm F}$ and $E_{\rm M}$}
\begin{tabular}{cccccccc}
\hline \hline
& $j_{\rm p}$ & $j_+$ &  $\phi_{\rm M}$ & $j-j_0$ & $\alpha_\phi$
& $|E_{\rm F}|$ & $|E_{\rm M}|$  \\ \hline
Case 1 & 0.30 & 0.25 & 1000 & 0.55 & $3.2 \times 10^{-4}$ & 18.5 & 25.7 \\
Case 2 & 0.30 & 0.25 & 600 & 0.55 & $1.6 \times 10^{-3}$ & 21.2 & 29.3 \\
Case 3 & 0.90 & 0.70 & 1000 & 1.60 & $5.4 \times 10^{-4}$ & 32.9 & 46.2 \\
Case 4 & 0.90 & 0.70 & 600 & 1.60 & $2.7 \times 10^{-3}$ & 36.8 & 52.0 \\ \hline
\end{tabular}
\end{minipage}
\end{table*}
%%%%%%%%%%%%%%%%%%%%%%%%%%%%%%%%%%%%%%%%%%%%%%%%%%%%%%%%%%%%%%

Region I ($l \le -800$) is not shown in Fig. 4.
The effective charge density is already positive
at the left edge ($l=-50$) of this figure.
The density of positrons initially
grows outside the PPF, but begins to decrease around the point
where the positrons injected at the PPF turn around.
Although the total charge density decreases once from this point,
the charge density of protons immediately makes the total charge density
increase again.
In regions III and IV the densities of both pair-electrons and positrons
increase while the charge density due to pairs is maintaining, $\bar{\rho}_+
+ \bar{\rho}_- \simeq -j_+$.
The potential of the valley for Case 1 is $\sim 600$ and high enough.

In our examples $\alpha_\phi$ turns out to be $10^{-4}$-$10^{-3}$,
which is much smaller than the value required in SMT02.
In our model the number of electron-positron pairs a beam-electron creates
per unit time at the PPF is $E_{\rm F} \alpha_\phi \omega_{\rm p}$,
which becomes $\sim 2 \times 10^8 T_{0.3}^{1/2}
B_{12}^{-1/2}$ ${\rm s^{-1}}$ for Case 1.
On the other hand, the photon number a primary beam electron emits per unit time
via curvature radiation is $2 e^2 \Gamma/3 r \hbar \simeq 10^7 \Gamma_6
(r/10^7 {\rm cm})^{-1}$ ${\rm s^{-1}}$, where $r$ is the radius of curvature
of the field line.
The resultant number of pairs created from these photons
will be of the same order.
Although we have adopted a very simple model and estimate the numbers roughly,
these numbers are rather close.
The required pair creation rate in our examples may be
consistent with the conventional pair creation models.

In order to keep the valley shallow enough,
our model requires that the negative charge due to pairs
flows into region IV, which leads to the negative charge density
of $-j_+$.
Since this charge density is very large compared with $\rho_{\rm in}$,
the total effective charge density remains negative in the valley.
Therefore, outside the valley the potential grows again.
Since the charge density due to protons is a function of only $\phi$,
this charge density dominates again as the potential grows,
and the potential again begins to decrease at $\phi=\phi_{\rm M}$ again.
Electron-positron pairs injected after the peak
flow outwards almost at light speed
(the lowest $\gamma_-$ at the valley
is $\sim 100$ for Case 1), so that these pairs do not contribute to
the charge density.
Reiterating this process, a space charge density wave
continues outside the valley.
This wave is essentially the same as the wave observed in the numerical solutions
in SMT98.
Of course, the average effective charge density is zero in this region.

\section{CONCLUSIONS AND DISCUSSION}

We have shown a possibility that the proton counterflow and injected electron-positron
pairs screen a strong electric field expected in the polar cap model.
In our numerical examples the required pair creation rate is consistent with
the prediction of the conventional pair creation models.
A space charge density wave appears outside the PPF in our model.
In order to screen a stronger electric field,
a larger proton current ($j_{\rm p} \simeq j-j_+-j_0$) and
a smaller potential difference between the PPF and the peak, $\phi_{\rm M}$,
are favorable.
The values of $\phi_{\rm M}$ and the Lorentz factor of pairs at injection
$\bar{\gamma}$ should be the same order.

In addition,
the amplitude of the charge density wave should be small enough to
keep pair-electrons from returning from outside the screening point.
The currents of returning positrons and protons should be comparable,
in order to achieve both a small wave amplitude and screening
of a large electric field.
Since a smaller $\phi_{\rm M}$
requires a higher pair creation rate,
the parameter is limited to $\phi_{\rm M} \sim O(10^3)$.
As for the pair Lorentz factor, $\bar{\gamma} \simeq 500$ is the best choice.
We found that the non-dimensional electric field
our model can screen is a few times ten at most.
To attain the Lorentz factor of the primary electron-beam, $\Gamma \sim 10^6$,
such a value of the electric field requires $\rho_{\rm in}<0.01$,
which means that the screening height is greater than $10^4$ cm.

In our model the flux of positrons and protons falling onto the star
is a few times the GJ flux.
The particles falling onto the star heat up the polar caps.
The X-ray thermal luminosity from the neutron star investigated by
{\it ROSAT} \citep{bec97} does not indicate that there is
a significant number of particles falling onto the star.
Therefore, the fraction of the active region in which the electron beam
is accelerated in the whole polar cap region should be small enough in this model.

Our model predicts the appearance of a space charge density wave after the screening.
Particles flow forward while accelerating and decelerating in turn.
Though the spatial average of the effective charge density is zero,
the local charge density is finite.
In more realistic cases, the potential wave may not be steady,
but oscillate with time.
The stability of the potential structure may be determined by the global dynamics
in the magnetosphere, which is beyond the scope
of this paper.
However, the existence of the wave is interesting when
we consider the pulsar emission mechanism.
The wave may induce the charge bunching that causes coherent radio emission.

If most ions in the co-rotating magnetosphere are Fe irons,
the counterflow could be iron flow.
In this case the scenario depends on the degree of ionization.
If the mass and charge of the Fe irons are $m_{\rm Fe}$ and $q e$, respectively,
$\delta$ should be replaced by $\delta'=q (m_{\rm p}/m_{\rm Fe}) \delta $ in our model.
The smaller $q$ becomes, the larger the typical amplitude of 
the potential variation becomes.
Thus we need to adopt a larger value of $\bar{\gamma}$
to avoid electrons returning.
Even if Fe irons are fully ionized ($q=26$),
$\bar{\gamma}$ is required to be more than $\sim 1000$
in our numerical estimate,
while $E_{\rm F}$ becomes larger than in the proton model.

\begin{acknowledgements}

This work is  supported in part by a Grant-in-Aid for Scientific Research 
from Ministry of Education and Science (No.13440061, F.T.).
One of the authors (K.A.) was supported by the Japan Society for the Promotion of Science.

\end{acknowledgements}

\end{document}